\documentclass{PoS}
\usepackage{amsmath,amssymb,amsthm,bm}
\usepackage{ascmac}
\usepackage{here}
\usepackage[]{multicol}
\usepackage{subfigure}

\makeatletter

\newenvironment{figurehere}
  {\def\@captype{figure}}
  {}
\makeatother

\newcommand{\DWilsonF}{D_{\mathrm{WF}}+1}
\newcommand{\sign}{\mathrm{sign}}

\title{A construction of the Schr\"odinger Functional for M\"obius Domain Wall Fermions}

\ShortTitle{A construction of the Schr\"odinger Functional for M\"obius Domain Wall Fermions}

\author{\speaker{Yuko Murakami} \\%
Graduate school of Science, Hiroshima University, Higashi-Hiroshima, Japan \\
        E-mail: \email{m132626@hiroshima-u.ac.jp}}

\author{ Ken-Ichi Ishikawa\\
Graduate school of Science, Hiroshima University, Higashi-Hiroshima, Japan \\
        E-mail: \email{ishikawa@theo.phys.sci.hiroshima-u.ac.jp}}

\abstract{
 {\normalsize        %
 \vspace*{-33.5em}   %
 \begin{flushright}  
 \ \ \ \ HUPD-1403   %
 \end{flushright}    %
 \vspace*{31em}\     %
 }                   %
We construct  the Schr\"odinger Functional (SF) setup for the M\"obius domain wall fermions (MDWF).
The method is an extension of the method proposed by Takeda for the standard domain wall fermion.
In order to fulfill the requirement that the lattice Dirac operator with the SF 
boundary obeys the L\"uscher's universality argument: the lattice chiral 
fermion with the SF boundary condition breaks the chiral symmetry at the temporal boundary,
we impose the parity symmetry with respect to the fifth-direction on the MDWF operator.
This additional symmetry restricts the choice of the parameter of the MDWF so that
the optimal parameter from the Zolotarev optimal approximation cannot be applied.
We introduce a modified parameter set having the fifth-dimensional parity symmetry. 
We investigate the MDWF with the SF boundary by observing eigenvalues of the Hermitian operator and
the Ginsparg-Wilson relation violation at the tree-level. 
We compare the computational cost with that of the standard DWF with the SF scheme.
}

\FullConference{The 32nd International Symposium on Lattice Field Theory\\
		23-28 June, 2014\\
		Columbia University New York, NY}

\begin{document}

\section{Introduction}
The lattice chiral symmetry with the Ginsparg Wilson (GW) relation~\cite{Ginsparg:1981bj} can be realized by 
the domain wall type or overlap type fermions, and the large scale simulations with these actions has 
been made to investigate QCD and flavor physics\cite{Aoki:2013ldr}. 
The renormalization factors for these actions are desirable and the Schr\"odinger functional (SF)
scheme~\cite{Luscher:1985iu} is one of the method and has been successfully used to investigate 
the running coupling constants,
running masses, and various renormalization factors non-perturbatively on the lattice.
However the realization of the lattice chiral symmetry with the SF boundary condition is not trivial 
because the SF temporal boundary condition must break the chiral symmetry at the temporal boundary 
in the continuum theory (the universality) as pointed by L\"{u}scher~\cite{Luscher:2006df};
\begin{align}
    \gamma_5 S(x,y)+S(x,y)\gamma_5 = 
\int_{z_0=0}d^3\bm{z}S(x,y)\gamma_5 P_{-}S(z,y)+
\int_{z_0=T}d^3\bm{z}S(x,y)\gamma_5 P_{+}S(z,y),
\label{eq:ChiralSymmProp}
\end{align}
where $S(x,y)$ is the massless Dirac propagator, $T$ is the temporal extent, and $P_{\pm}=(1\pm \gamma_4)/2$.
The lattice chiral fermions with the SF boundary condition should reproduce this relation 
in the continuum limit and have been constructed for the overlap
fermion~\cite{Luscher:2006df, Taniguchi:2004gf} and 
the standard domain wall fermion (SDWF)~\cite{Taniguchi:2006qw,Takeda:2009sk}. 
The overlap fermion with the SF scheme has been applied to the Gross-Neveu model~\cite{Leder:2007nj}.
The SDWF can be generalized by introducing the parameters which have the dependence on 
the index of the fifth-dimension to improve the chirality at a finite extent
in the fifth-direction~\cite{Chiu:2002ir, Brower:2004xi}. The SF construction of these generalized domain wall fermions 
are not known.
In this paper, we apply the SF boundary condition to the M\"{o}bius domain wall fermion~\cite{Brower:2004xi} (MDWF) 
aiming for constructing the SF scheme with the lattice chiral symmetry more effectively.

In the next section, we briefly introduce the SF construction for 
the SDWF and the boundary operator, which is designed to satisfy Eq.~(\ref{eq:ChiralSymmProp}), introduced 
by Takeda~\cite{Takeda:2009sk}.  
Then we apply them to the MDWF operator to break properly the chiral symmetry 
at the SF boundary. In this extension we need the fifth-direction parity for the MDWF. 
In section~\ref{sec:Sec3}, 
we introduce MDWF parameters into this operator to have the fifth-direction parity symmetry.
In section~\ref{sec:Sec4}, we check the universality of the MDWF operator with the SF boundary 
by investigating the spectrum and the chiral symmetry towards the continuum limit at the tree-level and 
we summarize this paper in the last section.

\section{A construction of the MDWF with the SF boundary condition}
\label{sec:Sec2}

The SDWF operator with the SF boundary term~\cite{Takeda:2009sk} ,$D_{\mathrm{DWF}}^{\mathrm{SF}}$, is 
\begin{align}
 &D_{\mathrm{DWF}}^{\mathrm{SF}}(n,s_5;m,t_5)  = (D_{\mathrm{DWF}} + B_{\mathrm{SF}})(n,s_5;m,t_5)\notag\\
&=
  \begin{pmatrix}
                  \DWilsonF & -P_L              &  0                    &                    0  &               0  & m_f P_R + c_{\mathrm{SF}}B  \\
                       -P_R & \DWilsonF         & -P_L                  &                    0  & c_{\mathrm{SF}}B & 0  \\
                         0  & -P_R              & \DWilsonF             & -P_L+c_{\mathrm{SF}}B &               0  & 0  \\
                         0  &               0   & -P_R-c_{\mathrm{SF}}B & \DWilsonF             & -P_L             & 0  \\
                         0  & -c_{\mathrm{SF}}B &                    0  & -P_R                  & \DWilsonF        & -P_L \\
 m_f P_L -c_{\mathrm{SF}} B &               0   &                    0  &                    0  & -P_R             & \DWilsonF \\
  \end{pmatrix}(n;m),
\end{align}
where
 $D_{\mathrm{DWF}}$ is the SDWF operator, 
  $D_{\mathrm{WF}}$ is the four dimensional Wilson-Dirac fermion operator with a negative mass,
$P_{R/L}=(1\pm \gamma_5)/2$, $c_{\mathrm{SF}}$ is the boundary coefficient, and $m_f$ is the mass parameter. 
The temporal hopping connecting the sites with the temporal site index $n_4=0$ and $T$ are zero in $D_{\mathrm{WF}}$ as usual with the SF boundary condition.
The boundary operator $B_{\mathrm{SF}}$ is defined by
\begin{align}
 B_{\mathrm{SF}}(n,s_5;m,t_5) &= c_{\mathrm{SF}}  f(s_5) B(n;m) \delta_{s_5, N_5 -t_5+1}, \\
 B(n,m) &= \delta_{\bm{n},\bm{m}} \delta_{n_4,m_4} \gamma_5
                (\delta_{n_4,1} P_L + \delta_{n_4,T-1} P_R), \\
 f(s_5) &=
\begin{cases}
 +1 & (1 \leq s_5 \leq N_5/2 ) \\
 -1 & (N_5/2 +1 \leq s_5 \leq N_5).
\end{cases}
\end{align}
In the following we restrict our attention  to the case $N_5$ with an even number and use $N_5=6$ as an example for this paper.
The structure of $B_{\mathrm{SF}}$ is almost uniquely fixed by the discrete symmetries ($C$, $P$, $T$, $\Gamma_5$-Hermiticity)
and the chiral symmetry breaking property at the boundary~\cite{Luscher:2006df, Takeda:2009sk}.

\newcommand{\DWFA}{D_{\mathrm{WF}}}
\newcommand{\DP}{D^{+}}
\newcommand{\DM}{D^{-}}

The MDWF operator is a generalization of the DWF operator 
aiming for better chiral property and cost-effectiveness~\cite{Brower:2004xi}. 
The MDWF includes the SDWF, Borici's DWF~\cite{Borici:1999da} and Chiu's optimal DWF~\cite{Chiu:2002ir} as the special cases.

We introduce the following operator as the MDWF operator with the SF boundary term $B_{\mathrm{SF}}$.
{\small
\begin{align}
 &D_{\mathrm{MDWF}}^{\mathrm{SF}}(n,s_5;m,t_5)  = (D_{\mathrm{MDWF}} - D_{\mathrm{D}}^{-}B_{\mathrm{SF}})(n,s_5;m,t_5)
\notag\\
& =
  \begin{pmatrix}
                                \DP_1 & \DM_1 P_L              &  0                                &                                  0  &                       0  & \DM_1 (-m_f P_R - c_{\mathrm{SF}} B)  \\
                            \DM_2 P_R & \DP_2                  & \DM_2 P_L                         &                                  0  & - c_{\mathrm{SF}}\DM_2 B & 0  \\
                                   0  & \DM_3 P_R              & \DP_3                             & \DM_3 (P_L - c_{\mathrm{SF}} B) &                       0  & 0  \\
                                   0  &                    0   & \DM_3 (P_R +c_{\mathrm{SF}} B) & \DP_3                               & \DM_3 P_L                & 0  \\
                                   0  & c_{\mathrm{SF}}\DM_2 B &                                0  & \DM_2 P_R                           & \DP_2                    & \DM_2 P_L \\
 \DM_1 (-m_f P_L +c_{\mathrm{SF}} B) &                    0   &                                0  &                                  0  & \DM_1 P_R                & \DP_1 \\
  \end{pmatrix}
  (n;m),
  \label{eq:MDWFDEF}
\end{align}
\begin{align}
& D^{-}_{\mathrm{D}}  = \mathrm{diag}(D^{-}_1,D^{-}_2,D^{-}_3,D^{-}_3,D^{-}_2,D^{-}_1),\\
& D^{+}_i = D_{\mathrm{WF}} b_i + 1, \quad\mbox{and}\quad
  D^{-}_i = D_{\mathrm{WF}} c_i - 1, \quad \mbox{($i=1,2,\cdots,N_5/2$)}.
  \label{eq:MDWFDEF2}
\end{align}
}
The tunable parameters $b_i$ and $c_i$ have the parity symmetry so that the MDWF operator satisfies 
the discrete symmetries $C,P,T,\Gamma_5$.
Because of this symmetry we cannot apply the optimal choice for the parameters, for example,
the optimal choice via the Zolotarev approximation introduced by Chiu~\cite{Chiu:2002eh} does not have this symmetry.
This parity symmetry is not the required condition for the usual temporal boundary condition (periodic/anti-periodic), nevertheless 
it seems to have theoretical benefits to analyze the operator and the action~\cite{Brower:2004xi,Chen:2014hyy,Ogawa:2009ex,PBOYLELat15, Furman:1994ky}
.
In order to improve the chiral property of the MDWF operator in accordance with the SF boundary condition,
we have to search an optimal choice for the coefficients $b_i$ and $c_i$ under 
the restriction of the parity symmetry.

\section{The quasi optimal Zolotarev approximation}
\label{sec:Sec3}
In the previous section, we have introduced the fifth-direction parity symmetry to the MDWF. 
An optimal choice for the coefficients $b_i$ and $c_i$ with the parity symmetry has been proposed and used
to construct the single flavor algorithm for the optimal domain wall fermion~\cite{Chen:2014hyy,Ogawa:2009ex}. 
In this section we briefly discuss our choice for the coefficients $b_i$ and $c_i$, and describe the property.

Without the SF boundary operator, the MDWF operator induces the following truncated overlap operator \cite{Borici:1998mr, Borici:2004pn, Kikukawa:1999sy};
\begin{align}
    D^{(N_5)}_{\mathrm{EOVF}} &\equiv 
 \epsilon^\dag P^\dag \left.D_{\mathrm{MDWF}}^{-1}\right|_{m_f=1} D_{\mathrm{MDWF}} P \epsilon
=\dfrac{1+m_f}{2} + \dfrac{1-m_f}{2}\gamma_5 R_{N_5}({\cal H}_W) , \\
 P &= P_L \delta_{s_5,t_5} + P_R \delta_{s_5,t_5+1} + P_R \delta_{N_5,1},
\end{align}
where $\epsilon=(1,0,0,0,0,0)^{t}$ is the operator projecting out a four dimensional slice 
and $P$ is the permutation matrix.
The matrix function $R_{N_5}(x)$ and the kernel operator ${\cal H}_W$ become
\begin{align}
    R_{N_5}(x) &= 
\dfrac{\prod_{j=1}^{N_5}(1+\omega_j x)-\prod_{j=1}^{N_5}(1-\omega_j x)}
{\prod_{j=1}^{N_5}(1+\omega_j x)+\prod_{j=1}^{N_5}(1-\omega_j x)},
\label{eq:RationalApprox1}
\\
{\cal H}_W & = \gamma_5 D_{\mathrm{WF}} (\alpha D_{\mathrm{WF}}+2)^{-1},\\
\alpha  & = b_j - c_j, \quad \omega_j = (b_j + c_j).
\end{align}
$\alpha$ is the M\"{o}bius parameter.
This converges to the sign function with appropriate conditions on $\omega_j$ and on the spectrum of ${\cal H}_W$. 
The optimal approximation for the sign function has been given in~\cite{Chiu:2002ir,Chiu:2002eh} and the coefficient 
dose not have the parity symmetry.

Under the parity constraint, Eq.~(\ref{eq:RationalApprox1}) becomes
\begin{align}
\tilde{R}_{N_5}(x) &\equiv
\dfrac{\prod_{j=1}^{N_5/2}(1+\omega_j x)^2-\prod_{j=1}^{N_5/2}(1-\omega_j x)^2}
{\prod_{j=1}^{N_5/2}(1+\omega_j x)^2+\prod_{j=1}^{N_5/2}(1-\omega_j x)^2}.
\label{eq:RationalApprox2}
\end{align}
Our choice for $\omega_j$ is simply to employ $\omega_j$ obtained for the half order ($N_5/2$) Zolotarev optimal
approximation $R_{N_5/2}(x)$~\footnote{This choice may have been already used in \cite{Chen:2014hyy,Ogawa:2009ex} for the single flavor simulation.}.
This choice for $\omega_j$ (and $b_j$ and $c_j$) violates the mini-max optimal approximation to the sign function
even if $R_{N_5/2}$ is the optimal Zolotarev approximation.
However the approximation error stays at the same order to the optimal one as seen from the following error analysis.
Because $\tilde{R}_{N_5}(x)$ can be written in terms of $R_{N_5/2}(x)$ as
\begin{align}
\tilde{R}_{N_5}(x)= 2 R_{N_5/2}(x)/ \left[\left(R_{N_5/2}(x)\right)^2+1\right],
\end{align}
the approximation error is bounded by
\begin{align}
 | \sign(x) - \tilde{R}_{N_5}(x) | \leq \dfrac{(\Delta_{N_5/2})^2}{2(1-\Delta_{N_5/2})+(\Delta_{N_5/2})^2} \equiv \tilde{\Delta}_{N_5}, \; \mbox{with} \; \Delta_{N_5/2} \equiv |\sign(x)-R_{N_5/2}(x)|.
\end{align}
Since the empirical error estimate indicates that $\Delta_{N_5} \sim (\Delta_{N_5/2})^2$~\cite{Chiu:2002eh} 
for the Zolotarev optimal approximation, we conclude that $\tilde{\Delta}_{N_5} \sim \Delta_{N_5}$ (see Fig.~\ref{fig:error}).
This choice is not optimal under the parity constraint, nevertheless we refer this choice of the coefficient as the quasi 
optimal approximation.

We employ the quasi optimal coefficients for the MDWF with the SF boundary term. 
In the following we restrict the kernel operator to the Shamir type kernel in order to compare them 
with the SDWF;
\begin{align}
    b_j & = ({\omega_j+1})/{2},\quad c_j = ({\omega_j-1})/{2}\quad(j=1,\cdots,N_5/2),\\
 {\cal H}_W &= \gamma_5 D_{\mathrm{WF}}(D_{\mathrm{WF}}+2)^{-1}.
\end{align}
$\omega_j$ is adjusted optimally to enclose the spectrum of ${\cal H}_W$.
Although the ordering of $\omega_j$ is arbitrary under the parity symmetry constraint,
we employ the ordering of $\omega_1 < \omega_2 < \cdots < \omega_{N_5/2}$.
\vspace*{-0.5em}
\begin{multicols}{2}
\begin{figurehere}
  \begin{center}
   \includegraphics[scale=0.5, trim=0 30 0 -30]{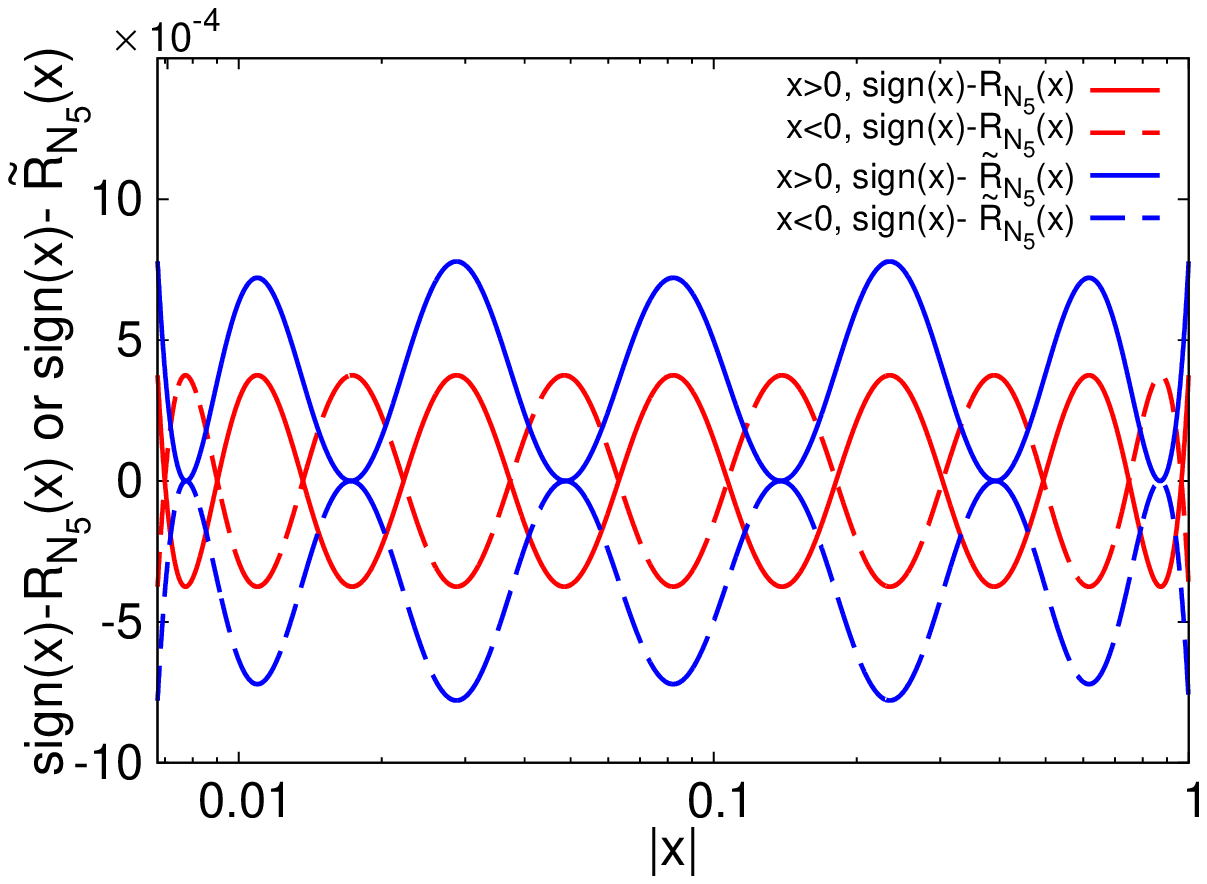}
  \end{center}
  \caption{The sign function approximation error for the optimal Zolotarev approximation (red) and the quasi optimal approximation (blue).
           The solid (dot) lines are error in the positive (negative) region of $x$.
		$N_5=12$, the approximation range is from $6.7497331 \times 10^{-3}$ to $1$.}
  \label{fig:error}
\end{figurehere}
\begin{figurehere}
  \begin{center}
   \includegraphics[scale=0.65, trim=30 10 0 0]{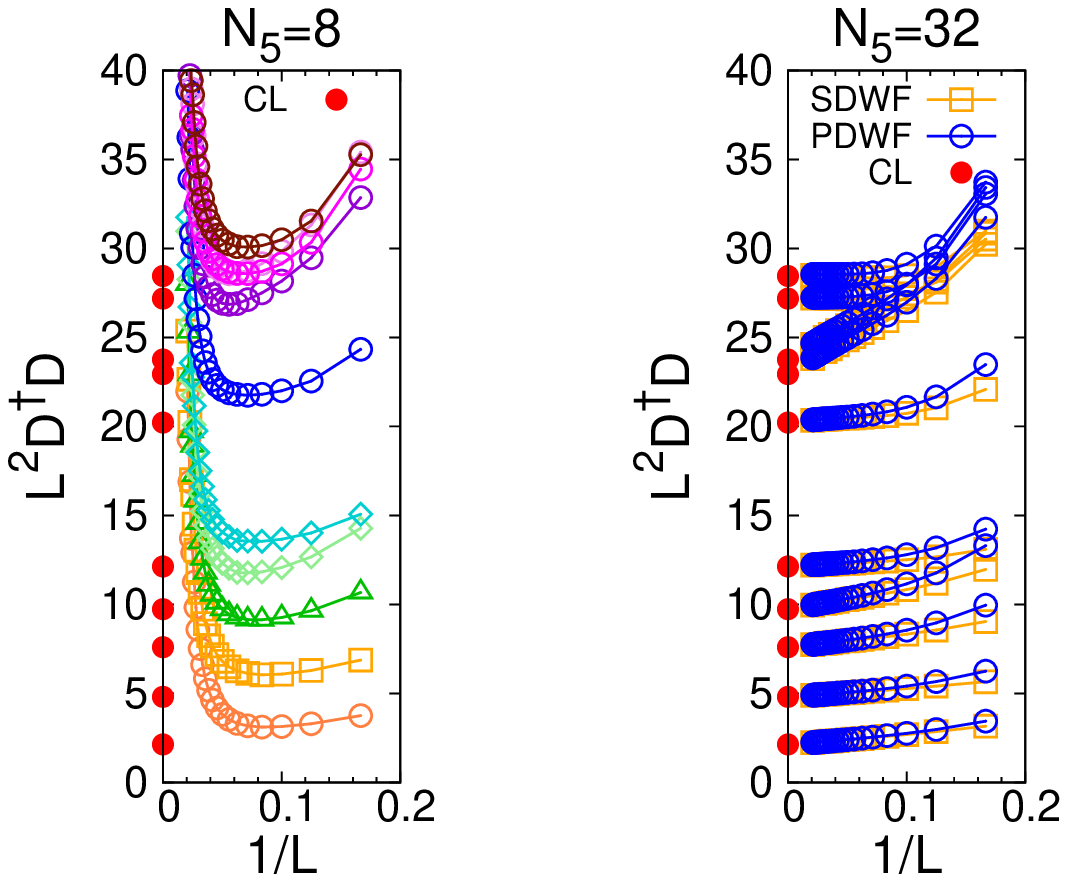}
    \caption{The lowest ten eigenvalues of the Hermitian operator, $N_5=8$ with PDWF (left) and $N_5=32$ with SDWF and PDWF (right).
              Red circles are the continuum values (CL) taken from~\cite{Sint:1995ch}.}
    \label{fig:hermite}
  \end{center}
\end{figurehere}
\end{multicols}

\section{The universality check}
\label{sec:Sec4}
We construct the Shamir optimal type DWF by applying the quasi optimal Zolotarev approximation coefficients, Eqs.~(\ref{eq:MDWFDEF})-(\ref{eq:MDWFDEF2}), and call this the Palindromic-optimal DWF (PDWF). 
We employ the standard boundary condition for the gauge field~\cite{Luscher:1993gh} which 
induces the classical background field, $m_f=0$, $M_0=1$ (negative mass parameter in $D_{\mathrm{WF}}$), $L=T$, and $c_{\mathrm{SF}}=1$ 
for both the PDWF and the SDWF.
In this section we study the property of the PDWF operator at the tree-level 
whether the operator satisfies the L\"{u}scher's universality argument using the double-precision arithmetic.

We investigate the lowest ten eigenvalues of the squared Hermitian operator $L^2 D_q^\dag D_q$, 
where $D_q$ is given by the following relation~\cite{Takeda:2009sk, Kikukawa:1999sy}, 
\begin{align}
{D_q}^{-1} \equiv (D^{(N_5)}_{\mathrm{EOVF}})^{-1}(1-D^{(N_5)}_{\mathrm{EOVF}}).
\end{align}
Figure~\ref{fig:hermite} shows the eigenvalues for the PDWF and the SDWF.
The red circles at $1/L=0$ are the eigenvalues in the continuum limit~\cite{Sint:1995ch}.
We find that the eigenvalues for the lattice fermion operator approach to those of the continuum operator 
appropriately when the lattice extent in the fifth-direction is large enough (right figure: $N_5=32$). 
When $N_5$ is small (left figure: $N_5=8$), however, the eigenvalues are leaving from the continuum values as decreasing $1/L$.
The reason is the following; the lowest eigenvalue of the kernel operator approaches to zero as decreasing $1/L$
and this makes the sign function approximation poor  with $N_5$ fixed at constant.
The continuum limit is properly realized as expected when the accuracy of the sign function approximation 
is good enough and the parameters are properly renormalized on a fixed constant physics.
The comparison between the SDWF and PDWF at $N_5=32$ shows that 
the lattice spacing error for the PDWF is slightly larger than that of the SDWF.

\begin{figure}[H]
	\begin{center}
	\includegraphics[scale=0.96, trim=0 75 0 75]{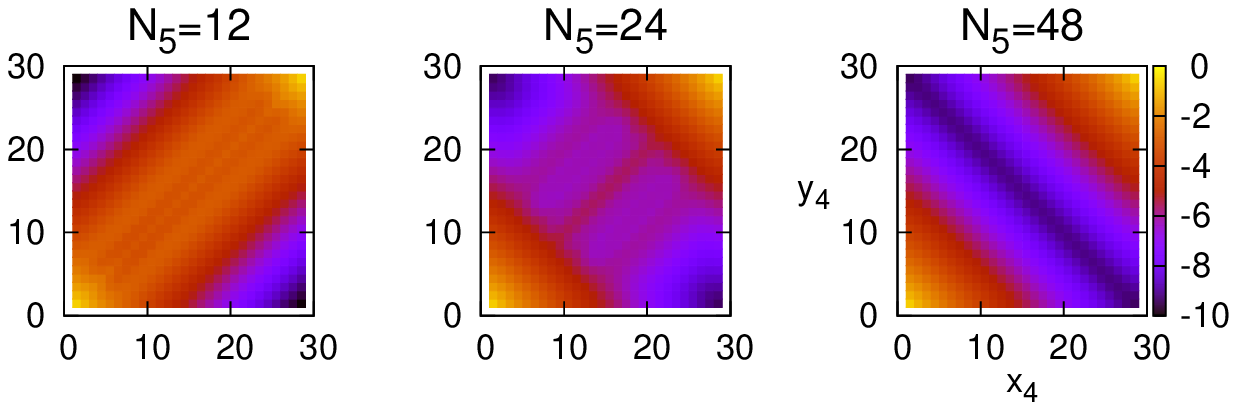}
	\caption{The time dependence of the GW relation violation of the PDWF operator.}
	\label{fig:GW}
	\end{center}
\vspace*{-1em}
\end{figure}

\begin{figure}[H]
	\begin{center}
	\includegraphics[scale=.5, trim=0 20 0 30]{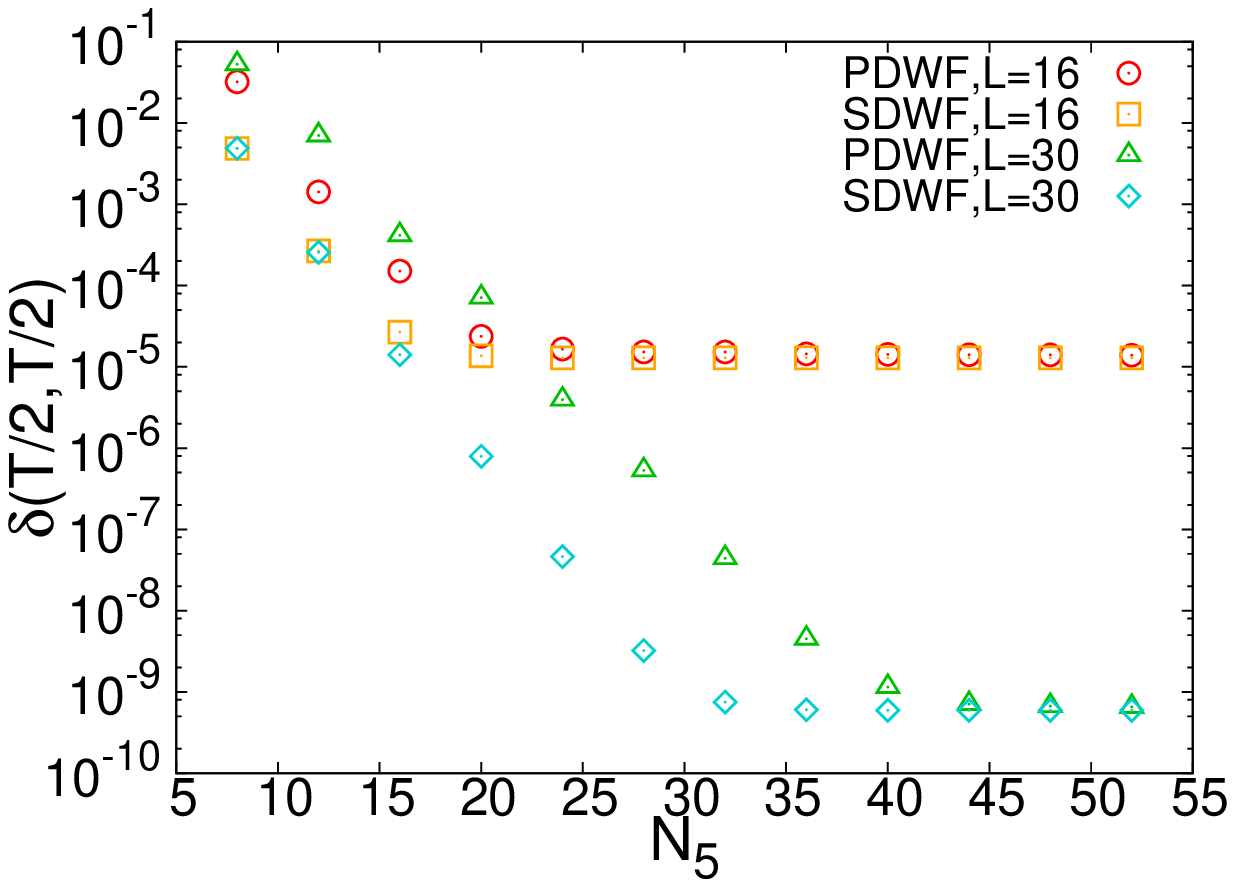}
	\caption{The GW relation violation as a function of $N_5$.}
	\label{fig:comp_gw}
	\end{center}
\vspace*{-1em}
\end{figure}

In order to see the chiral symmetry violation effect of the boundary term $B_{SF}$ we examine 
the GW relation violation in the temporal direction $\delta_{GW}(n_4,m_4)$;
{\small
\begin{align}
\delta_{GW}(n_4,m_4) = \max_{\mathrm{color}} \left|
\gamma_5 D^{(N_5)}_{\mathrm{EOVF}}(\bm{p},n_4;m_4)
+ D^{(N_5)}_{\mathrm{EOVF}}(\bm{p},n_4;m_4)\gamma_5
- 2 (D^{(N_5)}_{\mathrm{EOVF}}\gamma_5D^{(N_5)}_{\mathrm{EOVF}}) (\bm{p},n_4;m_4) \right|
,
\end{align}}
with spatial momenta $\bm{p}=0$. 
Figure~\ref{fig:GW} shows the time dependence of $\delta_{GW}(n_4,m_4)$ 
with $L=T=30$ in common logarithmic scale.
We observe that the chiral symmetry in the bulk region is restored as increasing $N_5$, while
the chiral symmetry violation remains only at the SF temporal boundaries. 
Although this does not reflect Eq.~(\ref{eq:ChiralSymmProp}) directly, this is desired behavior 
for the universality argument.
We show that the GW relation violation at the center of the temporal lattice for the SDWF and the PDWF 
with $L=T=16$ and $=30$ in Figure~\ref{fig:comp_gw}.
The violation decreases as increasing $N_5$ and is bounded from below and the error bound becomes 
smaller as decreasing the lattice spacing.
The error bound seems to be the finite lattice spacing error induced by the SF boundary condition.
As seen in  Figure \ref{fig:comp_gw}, the SDWF has a smaller error than that of the PDWF 
at the same $N_5$ before reaching the error bound.
The PDWF is not cost-effective. This is unexpected and we need further investigation.


\section{Summary}
\label{sec:Sec5}
We have constructed the M\"{o}bius domain wall fermion (MDWF) with the SF boundary condition in this paper.
In order to introduce the proper boundary condition and the desired property on the operator 
we imposed the parity symmetry in the fifth-direction on the MDWF operator with the SF boundary term.
We have introduced the quasi optimal Zolotarev approximation which satisfies the parity symmetry
and constructed the Palindromic-optimal domain wall fermion (PDWF) operator with the SF boundary term.
We investigated the lower eigenvalues and the GW relation violation of the PDWF operator 
and compared them to those of the standard DWF operator.
The continuum limit of the spectrum was properly recovered and the desired chiral symmetry property
were observed at the tree-level analysis. 
However the quasi optimal approximation does not improve the chiral symmetry.
One reason of this behavior could be the effect of the $O(a)$-error coming from the boundary term. 
This error can be removed by tuning $c_{\mathrm{SF}}$.
We have to investigate $c_{\mathrm{SF}}$ and the universality of the beta function at the one-loop level,
and these are ongoing.

\section*{Acknowledgements}
We would like to thank S.~Takeda for the computational advice and the useful discussion.
This work was supported in part
by a Grant-in-Aid for Scientific Research (C) (No. 24540276) from the Japan Society for the
Promotion of Science (JSPS).


\end{document}